# Ensemble CNNs for Breast Tumor Classification

Muhammad Umar Farooq
Department of IT Energy Convergence, Korea National University of Transportation, South Korea
umarfarooq@a.ut.ac.kr

Zahid Ullah
Department of Software, Korea National University of Transportation, South Korea
zeeuom@gmail.com

Jeonghwan Gwak
Department of Software, Korea National University of Transportation, South Korea
james.han.gwak@gmail.com

## ABSTRACT

To improve the recognition ability of computer-aided breast mass classification among mammographic images, in this work we explore the state-of-the-art classification networks to develop an ensemble mechanism. First, the regions of interests (ROIs) are obtained from the original dataset and then three models, i.e., XceptionNet, DenseNet, and EfficientNet, are trained individually. After training, we ensemble mechanism by summing the probabilities outputted from each network which enhances the performance up to 5%. The scheme has been validated on a public dataset and we achieved accuracy, precision, and recall 88%, 85%, and 76% respectively.

## 1 INTRODUCTION

Breast cancer is the most common type of cancer in women which causes approximately 40,000 deaths each year in the U.S. [1]. The death rate can be reduced significantly by identifying the breast cancer in the early stage. Using mammographic images the detection of breast cancer is a cost effective technique, and analyzing these images, the radiologists can make a diagnosis. However, the day by day production of large number of mammographic images brings a huge workload on radiologists and it also leads to misdiagnosis. Therefore, the development of an automated computer-aided diagnosis system is required that can improve the diagnosis accuracy and relieve the pressure on radiologists. The computer aided diagnosis system can assist the radiologists in distinguishing the normal and abnormal tissues. The automatic diagnostic system for mammographic images initially extracts the region of interest and then classify the region of interest into benign and malignant tissue. Also training these models needs a huge amount of annotations, including segmentation ground truths or bounding boxes in the training set. Unfortunately, experienced radiologists are required for annotating mammograms with expert domain knowledge put a significant amount of effort to ensure the accuracy, which will significantly maximize the workload of radiologists. The other key challenge in mammographic image analysis is difference between mammographic images and RGB images, that makes it difficult to apply classification models with good performance on RGB images to mammographic images. In breasts, masses are typically isodense or dense, thus it has the characteristics of pixel intensity from gray to white. Geometrically, they can be round, oval or irregular in shape with circumscribed, spiculated, obscured or ill-defined margins [2].

Recently, more and more researchers have begun to study deep learning-based models, due to the excellent performance of deep learning models in the field of computer vision [3,4]. For instance, Wang et al [5], developed a deep learning model called stacked denoising auto-encoder to classify the malignant and benign lesions and obtained satisfactory results. Chougrad et al [6] proposed a mass lesion mammography classification system using CNN. During training phase, they used the fine-tuned and transfer learning approach. They evaluated their proposed model using DDSM and BCDR datasets and obtained reasonable accuracy.
Li et al. [7] proposed four classification models using the CNN technique such as, CNN-2d, CNN-4d, CNN-2, and CNN-4. Among these techniques, the CNN-4d perform well and obtained 89.05% accuracy, 90.63% sensitivity, and 87.67% specificity.

Neeraj Dhungel et al. [8] developed a deep learning based method that automatically segments the area of lesions and then classifies the mammogram. They achieved 0.74 ROC rate for whole image, 0.80 ROC rate for whole image with automatically detected small lesion patches, and 0.91 ROC rate for whole image plus manually segmented small patches. In general, the mammograms classification using small abnormality patches affords reasonable performance, however, it requires extensive pre-processing work.

## 2 PROPOSED METHOD

The proposed scheme exploits three state-of-the-art networks for the classification of breast tissues in mammogram into normal, benign, and malignant. After that, we apply an ensembled mechanism to further optimize the performance. The details of each network and ensembled scheme have been provided in succeeding subsections.

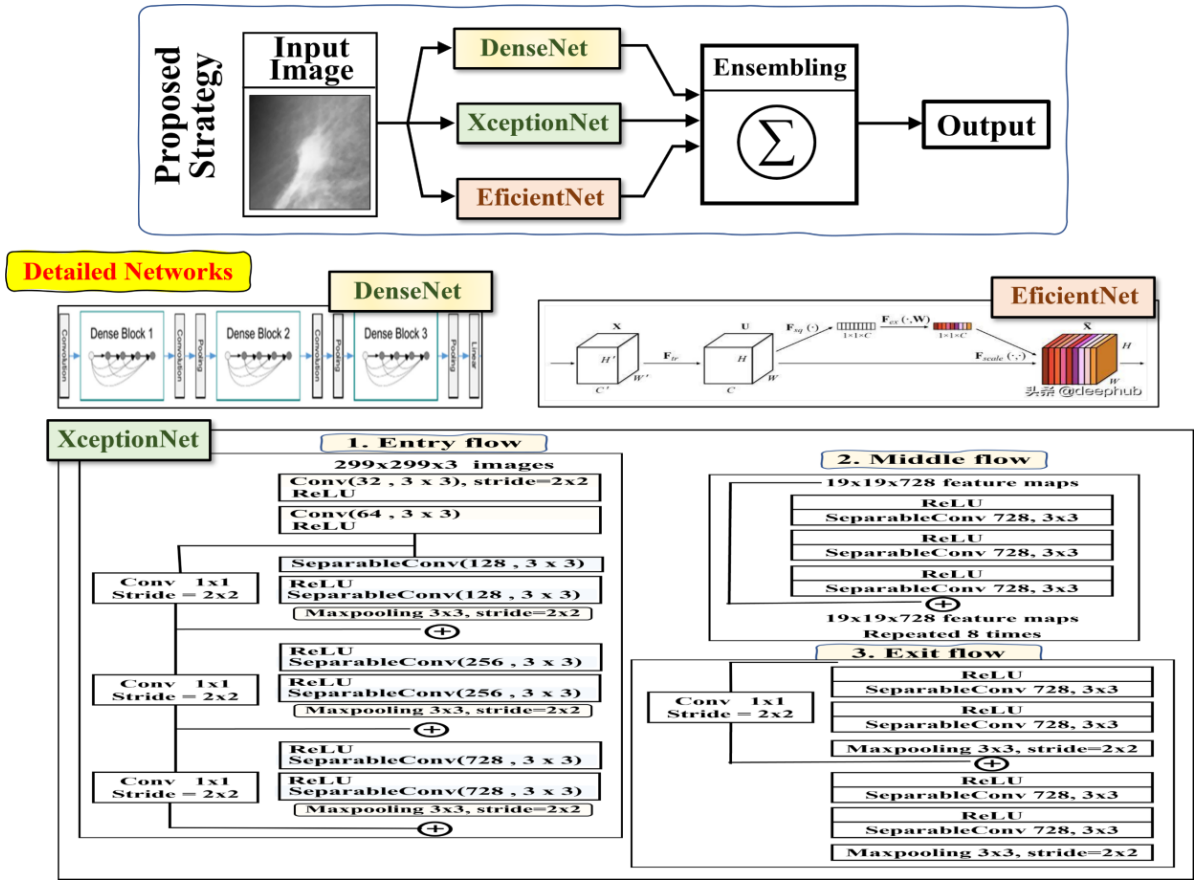

Figure 1: Our proposed model diagram.

## 2.1 EfficientNet

The EfficientNet model was proposed by Tan and Le, 2019 [9], which can achieve a suitable effect on the expansion of the depth, width, and resolution of the network, and then obtain good model performance. The EfficientNet Models are based on simple and highly effective compound scaling methods. This method enables to scale up a baseline ConvNet to any target resource constraints while maintaining model efficiency, used for transfer learning datasets. In general, EfficientNet models achieve both higher accuracy and better efficiency over existing CNNs such as AlexNet, ImageNet, GoogleNet, and MobileNetV2 [10]. EfficientNet could serve as a new foundation for future computer vision tasks. EfficientNet includes models from B0 to B7, and each one has different parameters from 5.3M to 66M.

## 2.2 DenseNet

DenseNets [11] has been introduced recently in literature. It reduces the connection between the input and output which helps in overcoming the vanishing gradient problem. Each layer in the DenseNet has a reduced feature map size which is important for training the CNN's on a small dataset leading to less probability of facing the over-fitting problems and to ensure that there is no loss in the transmitted information [41]. Additionally, each layer receives supervision from the loss function and a regularizing effect through shorter connections leading to an easier training process. The DenseNet is mainly composed of DenseBlock, Transition Layer, and Growth Rate.

## 2.3 XceptionNet

The Xception architecture introduced by Francois Chollet is an extension of the Inception architecture. This architecture is a linear stack of depthwise separable convolution layers with residual connections. The depthwise separable convolution aims to reduce computational cost and memory requirements. Xception has 36 convolutional layers structured into 14 modules, all include linear residual connections, except for the first and last modules. The separable convolution in Xception separates the learning of channel-wise and space-wise features. Also, the residual connection helps to solve the issues of vanishing gradients and representational bottlenecks by creating a shortcut in the sequential network. This shortcut connection is making the output of an earlier layer available as input to the later layer using a summation operation rather than being concatenated.

### 2.4 Ensembling Scheme

To ensemble three networks, we apply summation on the output probabilities and then choose the maximum as defined in the following Equation.

$$R = Max(Eff(x) + Dense(x) + XcepNet(x))$$

Where *Eff(x), Dense(x) and XcepNet(x)* are the probabilities for each class from EfficientNet, DenseNet and XceptionNet, respectively.

## 3 RESULTS AND DISCUSSION

The experiment results for Densenet121, Efficientnet_B4, XceptionNet, and ensembled scheme are shown in Table 1. The experiment is performed for 100 epochs with a learning rate of 0.001. From the table, the ensembled scheme achieves the highest accuracy compared to the three individual models.

To analyze the performance of each model separately and ensembled model, the confusion matrices are shown in Figure 2 which is based on, (a) DenseNet121, (b) EfficientNet_B4, (c) XceptionNet, and (d) is ensembled model performance respectively. The results demonstrate that the proposed ensemble scheme enhances the performance for the classification of breast tumor tissues.

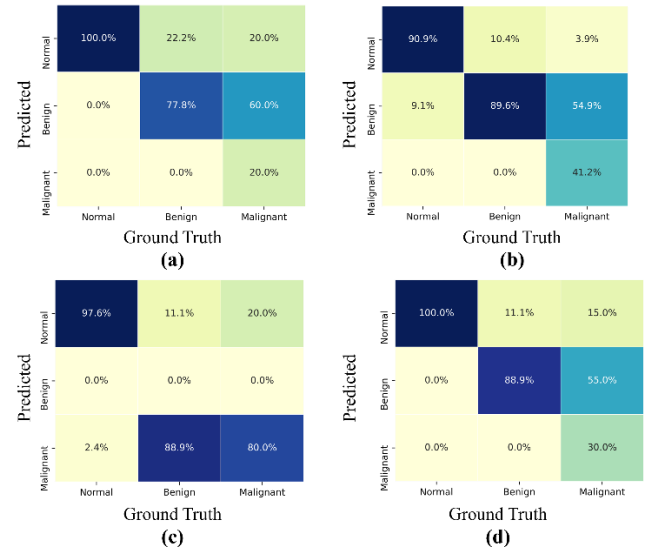

**Figure 2: Confusion matrix of deep learning models. (a) Densenet121. (b) Efficientnet_B4. (c) XceptionNet. (d) Ensembling scheme.**

## 4 CONCLUSION

This work presents an ensemble scheme for the classification of breast tissues in normal, benign, and malignant tissues, from mammograms. We use three state-of-the-art classification models (i.e., EfficientNet, DenseNet, and XceptionNet) with an ensemble mechanism. The core objective is to obtain the probability of each class from three classification models, and combine them to enhance the classification performance as compared to the individual obtained by each of them separately. The experiment results show that by using the ensembling scheme we are able to achieve the highest accuracy compared to the three individual model accuracies. The ensembled model's accuracy is 88.33% with precision, recall, and F1 scores of 85.62, 76.29, and 75.82 respectively. Future research plans include the exploration of more classification models and improve them further before ensembling to further optimize the results.

**Table 1: Experimental Results**

| Model | Accuracy | Precision | Recall | F1 Score |
|---|---|---|---|---|
| **DenseNet121** | 83.33 | 81.65 | 65.93 | 64.11 |
| **EfficientNet_B4** | 83.33 | 82.45 | 74.22 | 68.88 |
| **XceptionNet** | 80 | 46.69 | 59.19 | 51.49 |
| **Ensembling** | **88.33** | **85.62** | **76.29** | **75.82** |